# Self-similar motion for modeling anomalous diffusion and nonextensive statistical distributions

Zhifu Huang[1], Guozhen Su[1], Qiuping A Wang[2] and Jincan Chen[1,2,]*

[1)]Department of Physics and Institute of Theoretical Physics and Astrophysics, Xiamen University, Xiamen 361005, People's Republic of China

[2)]Institut Supérieur des Matériaux et Mécaniques Avancées du Mans, 44 Av. Bartholdi, 72000 Le Mans, France

We introduce a new universality class of one-dimensional iteration model giving rise to self-similar motion, in which the Feigenbaum constants are generalized as self-similar rates and can be predetermined. The curves of the mean-square displacement versus time generated here show that the motion is a kind of anomalous diffusion with the diffusion coefficient depending on the self-similar rates. In addition, it is found that the distribution of displacement agrees to a reliable precision with the q-Gaussian type distribution in some cases and bimodal distribution in some other cases. The results obtained show that the self-similar motion may be used to describe the anomalous diffusion and nonextensive statistical distributions.

___________________________

*Email: jcchen@xmu.edu.cn

Diffusion is one of the most important types of motion in nature. Its thorough understanding from dynamical and thermodynamic points of view is still a big matter of investigation. A widely investigated diffusion, among many others, is the anomalous one which occurs in many physical and biological systems [1-4] often having fractal and self-similar structures, long-range interaction and/or long-duration memory, and so on. Anomalous diffusion is characterized by one-dimensional mean-square displacement as follows:

$$<x^2(\Delta\tau)> \propto \Delta\tau^\alpha, \qquad (1)$$

where $x(\Delta\tau)$ is the displacement in time interval $\Delta\tau$, $\alpha$ is called the diffusion coefficient characterizing the time behavior of the mean-square displacement. The cases of $\alpha < 1$ and $\alpha > 1$ correspond to the subdiffusion and superdiffusion, respectively, while $\alpha = 1$ corresponds to the normal diffusion or Brownian motion.

The probability distribution of displacement in anomalous diffusion usually does not agree with Gaussian distribution coming from independent or nearly independent contributions, but may take the form of q-Gaussian given by

$$p(x) \propto [1-(1-q)\beta x^2]^{1/(1-q)}, \qquad (2)$$

where $\beta$ is a parameter characterizing the width of the distribution and q is the nonextensivity index [5-9]. Anomalous diffusion is sometimes associated with q-Gaussian distribution, as in the liquid with vortices [10] or in the driven-dissipative dusty plasma [11, 12], sometimes not [13]. In Eq.(2), $q \neq 1$ indicates a departure from the Gaussian shape while $q \to 1$ limit yields the normal Gaussian distribution.

As a family of stochastic motion, anomalous diffusion has been widely studied within various models and circumstances [14] such as fractional Brownian motion [15], Lévy motions [16], fractional stable Lévy motions [17, 18], to cite only some. However, in the analysis of stochastic process, the long-range interaction and/or long-duration memory can not be well described. On the other hand, the iterations of a deterministic dynamical system can never be completely independent from each other, since they are generated by deterministic algorithm. Concerning this matter, much work has been done to find the properties of iteration that allow a classification of deterministic systems. May [19] established a simple mathematical model called logistic map and founded the theory of complicated dynamics. It is shown that there exists periodic motion or chaos in a deterministic dynamical system, which enables us to analyze the statistical properties from deterministic dynamics by simple iterative algorithm. An intriguing aspect of the chaotic motion in complex systems usually exhibits self-similar structures [20] characterized by geometrical invariance under the change of scale and plays a central role in physics [21]. It is common knowledge that nonlinear maps are deterministic dynamics and often have self-similar structures. They exhibit various routes to chaos [22].

In particular, one-dimensional maps are of widely used tools to study the emergence of complexity in dynamical systems. Tirnakli et al. [23, 24] showed that the probability distribution of the sums of iterates at the edge of chaos of the z-logistic map is numerically consistent with a q-Gaussian distribution given by Eq.(2) with q=1.63. Moreover, Fergenbaum [25, 26] proved that the self-similar proportion is constant in all unimodal dissipative maps. Ruiz et al. [27] obtained several Feigenbaum-like constants in a new universality class of one-dimensional dissipative maps and derived several values of q by fitting the probability distributions to Eq. (2). Their iterations, however, can only yield one value of q each time and, therefore, can not give predetermined Feigenbaum-like constants. In addition, these dissipative maps can not meet the form of anomalous diffusion given by Eq.(1).

In this rapid communications, we attempt to construct a universality class of self-similar motion with a new method of iteration. The trajectory is iterated with self-similar structure in different scale in the whole period. Furthermore, we will analyze the properties of diffusion through simulating the underlying deterministic dynamical system.

In previous works, Cantor iterated a continuous line to get a self-similar line segment called the Cantor set. Similarly, let us attempt to use a simple iteration method to get a periodic motion which is similar to its own. Suppose that the periodic motion is of the square-wave form with the velocity $v_1$ in its front half period and $-v_2$ in its latter half period ($v_1, v_2 > 0$), i.e., the velocity in the first iteration is given by

$$v(1,t) = \begin{cases} v_1, & 0 \leq t < \dfrac{T}{2} \\ -v_2, & \dfrac{T}{2} \leq t < T \end{cases} \tag{3}$$

as shown in Fig. 1a, where $v_1$ and $v_2$ are equal to 1.2 and 0.9, respectively. In order to get a motion which is similar to its own in different time scale, we can iterate the front half period and latter half period similar to the whole period in the rate $r_1$ and $-r_2$, respectively, i.e.,

$$v(i+1,t) = \begin{cases} v_1 + r_1 v(i,2t), & 0 \leq t \leq \dfrac{T}{2} \\ -v_2 - r_2 v(i,2t), & \dfrac{T}{2} \leq t \leq T \end{cases} \quad i = 1,2,3.... \tag{4}$$

In order to get the motion similar to its original motion after iterated, the iteration rate is assumed to be proportional to the velocity, i.e., $\dfrac{r_1}{r_2} = \dfrac{v_1}{v_2}$. It can be seen from Eq.(4) that when the trajectory is iterated in the whole period, the self-similar rates of iteration $r_1$ and $r_2$ can be changed continuously and predetermined. It shows that the self-similar rates of iteration $r_1$ and $r_2$ in the present letter are of a generalization of the Feigenbaum constants [25, 26].

It is also seen from Eq.(4) that the number of states of the velocity will be doubled when the time interval of one state of the velocity is halved in one iteration. Thus, the time interval of one state of the velocity in the $i$ th iteration is

$$\Delta t(i) = \frac{T}{2^i}. \tag{5}$$

According to Eq.(5), when the number of iteration is large enough, the time interval of one state of the velocity will be far less than the period. In this situation, the number of states is large enough to analyze the statistical property of the motion. In the $i$ th iteration, the displacement in the time interval $n\Delta t(i)$ beginning from the stochastic time $t_{ran}$ can be numerically calculated by

$$x(i, t_{ran}, n) = \int_{t_{ran}}^{t_{ran} + n\Delta t(i)} [v(i,t) - \overline{v(i)}] dt, \tag{6}$$

where $n$ is a positive integer and $\overline{v(i)}$ is the average velocity in the iteration, which can be derived from Eq.(4) as

$$\overline{v(i)} = \frac{\frac{v_1 - v_2}{2}[1 - (\frac{r_1 - r_2}{2})^i]}{1 - \frac{r_1 - r_2}{2}}. \tag{7}$$

In order to get a reasonable statistical property, $n$ should be large enough, i.e., $n \gg 1$. In addition, the period should be far longer than the calculation time interval, i.e., $T \gg n\Delta t(i)$. Starting from Eq.(6), we first numerically calculate the displacement $x(i, t_{ran}, n)$ for each stochastic beginning time $t_{ran}$ and then find the mean square displacement as

$$\sigma^2(i,n) = <x^2(i, t_{ran}, n)>, \tag{8}$$

Each $\sigma^2(i,n)$ is the average of $x^2(i, t_{ran}, n)$ at different stochastic beginning times.

Using Eqs. (4)-(8), we can plot the curves of $\ln[\sigma^2(n)/\sigma^2(n_0)]$ varying with $\ln(n/n_0)$ for a very large $i$, as shown in Fig.2, where $n_0$ is equal to $2^{10}$. It is very interesting to note that if $n\Delta t$ is chosen to be equal to $\Delta \tau$ in Eq. (1), the slope of the curves in Fig.2 is just equal to $\alpha$ in Eq. (1). It illustrates clearly that the numerical simulations based on Eqs. (4)-(8) are in excellent agreement with Eq.(1) and the different choices of $r_1$ and $r_2$ correspond to the different values of the diffusion coefficient $\alpha$, respectively.

Using Eqs. (4)-(8), we can also plot the $r_1$ versus $r_2$ curve for $\alpha = 1$, as shown in Fig.3. It is obvious that the regions above and below the curve in the $r_1 \sim r_2$ plane correspond to the cases of subdiffusion ($\alpha < 1$) and superdiffusion ($\alpha > 1$), respectively. This indicates that the self-similar motion can well illustrate the anomalous diffusion in different cases and the cases of superdiffusion or subdiffusion can be well distinguished from different self-similar rates.

Moreover, we can do a statistical analysis on the displacements of self-similar motion mentioned above. Using Eqs. (4)-(8), we can obtain the distributions of the displacements of self-similar motion with different self-similar rates, as shown in Fig.4, where $r_1 = 1.3$ and $r_2 = 0.8r_1$, $1.0r_1$ and $2.0r_1$, respectively. At the same time, we can easily generate the q-Gaussian distribution curves for different values of q by using Eq. (2), as shown in the solid lines in Fig. 4. It is clearly seen from Fig. 4 that the distributions of the displacements of self-similar motion obtained by Eqs. (4)-(8) can be well fitted to the q-Gaussian distributions described by Eq. (2) and the different self-similar rates correspond to the different values of q, respectively. This demonstrates that not only the anomalous diffusion but also nonextensive statistical distributions can be well simulated by the self-similar motion. It is worth mentioning that, compared with the previous researches [23, 24, 27-29] where only single value or several discrete values of q were obtained, the present work is an important step forward in that the continuous values of q can be obtained about in the range of $1.1 < r_1 < 1.5$ and $0.8r_1 < r_2 < 2.0r_1$. The further investigation indicates that in some other ranges of $r_1$ and $r_2$, we can also obtain some other types of the distributions of the displacements of self-similar motion which are different from the q-Gaussian distributions. For example, when $r_1 = r_2 = 1.8$ or $r_1 = r_2 = 2.0$ is chosen, we obtain the bimodal distributions of the displacements of self-similar motion, as shown in Fig. 5. Such bimodal distributions have been found in the distributions of Epitaxial Island Growth [30] and neon nanobubbles in aluminum [31]. It shows that the bimodal distributions generated by complex systems may be simulated by self-similar motion as well.

To sum up, we have introduced a new model to generate a self-similar motion which is iterated in the whole period. It is found that the anomalous diffusion can be investigated by using self-similar motion with diffusion coefficients depending on self-similar rates. It is also found that self-similar motion can be directly used to analyze the q-Gaussian distributions and bimodal distributions appearing in nonextensive statistical mechanics and the different self-similar rates can reveal the different values of q. The results obtained here are helpful for the further understanding of the occurrence of anomalous diffusion, q-Gaussian and bimodal distributions in many natural, artificial, and social complex systems, and for the correct interpretation of experimental results in certain complex dynamical systems, in particular, in the ubiquitous dissipative systems. It is expected that the further research in this direction may open new perspectives.

**Acknowledgments**

This work has been supported by the National Natural Science Foundation (No. 10875100), People's Republic of China.

Figure captions:

Fig.1. The illustration of a self-similar motion for the parameters $r_1 = 1.2$ and $r_2 = 0.9$, where maps a, b, c and d correspond to the cases of i=1, 2, 3, 10, respectively.

Fig.2. The mean square displacement versus time interval curves, where $\sigma^2(n)$ represents the mean square displacement in the time interval of $n\Delta t$ and $n_0$ is equal to $2^{10}$ in simulation.

Fig.3. The $r_1$ versus $r_2$ curve for the diffusion coefficient $\alpha = 1$. The curve divides the $r_1 \sim r_2$ plane into two regions, which correspond to the cases of subdiffusion ($\alpha < 1$) and superdiffusion ($\alpha > 1$), respectively.

Fig.4. The distributions of displacements of self-similar motion for different self-similar rates, where $r_1 = 1.3$, $r_2 = 0.8r_1$, $1.0r_1$ and $2.0r_1$, respectively. The standard Gaussian curve and q-Gaussian curves with $q$=0.78, 1.36, and 1.96 are represented by dashed and solid lines, respectively.

Fig.5. The bimodal distributions of displacements of self-similar motion in different self-similar rates, where $r_1 = r_2 = 1.6$, $r_1 = r_2 = 1.8$, and $r_1 = r_2 = 2.0$ are chosen, respectively.

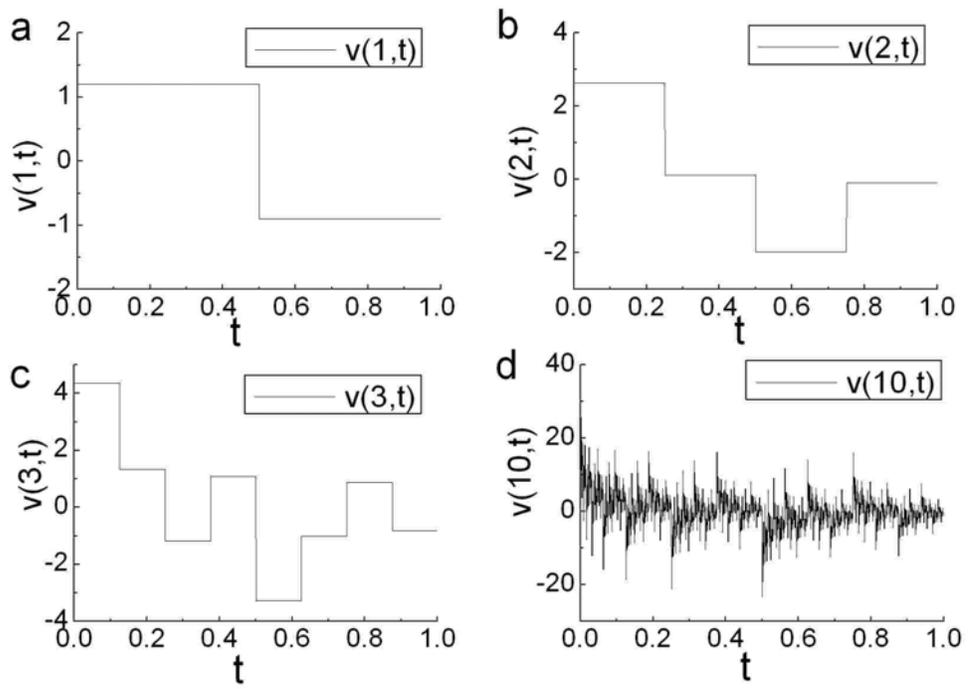

Fig.1.

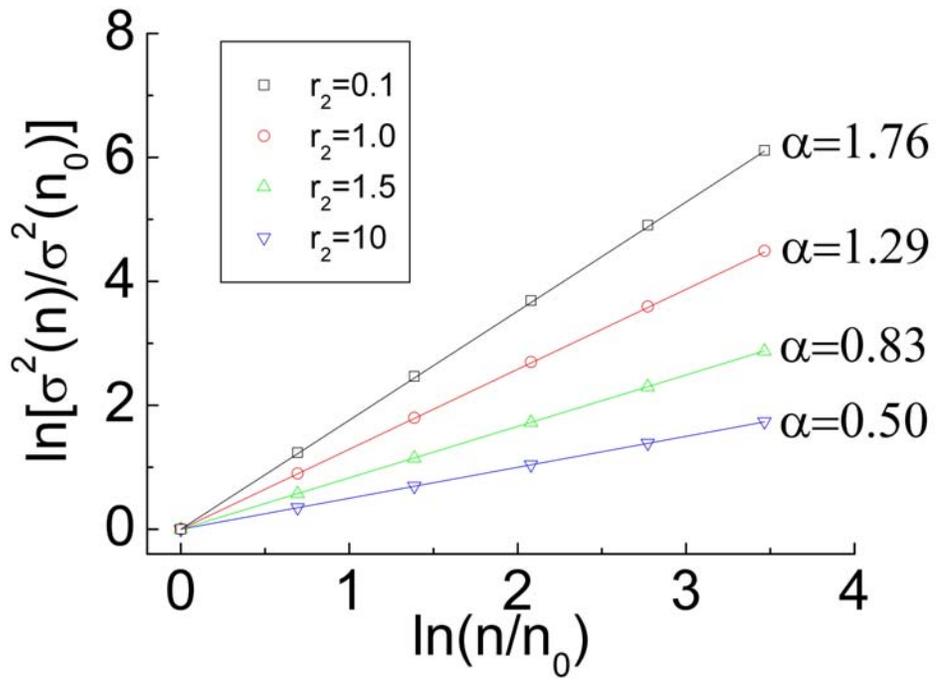

Fig.2.

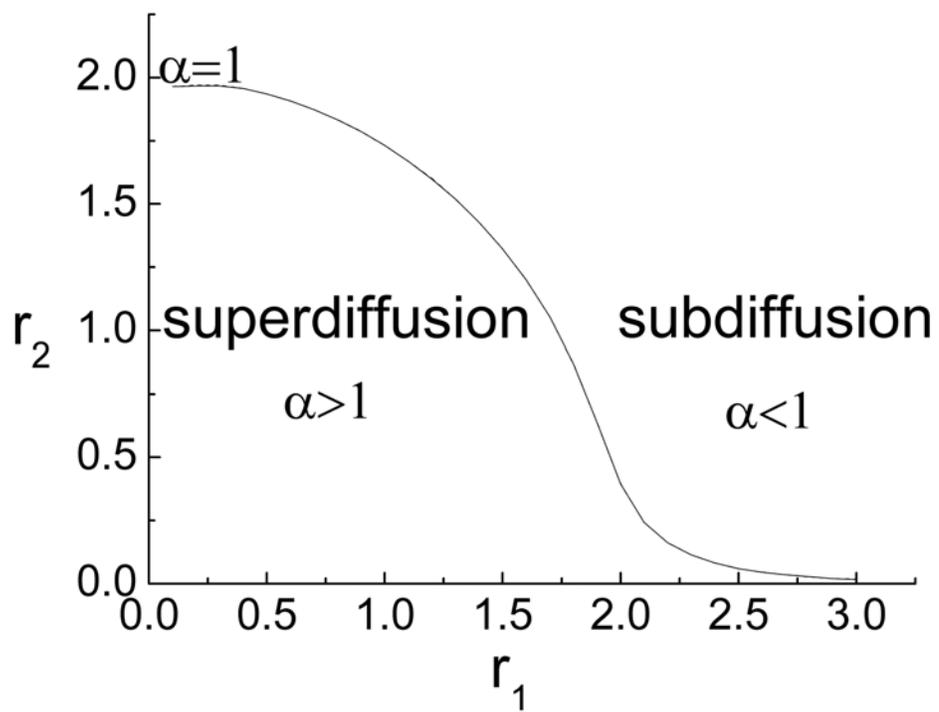

Fig.3

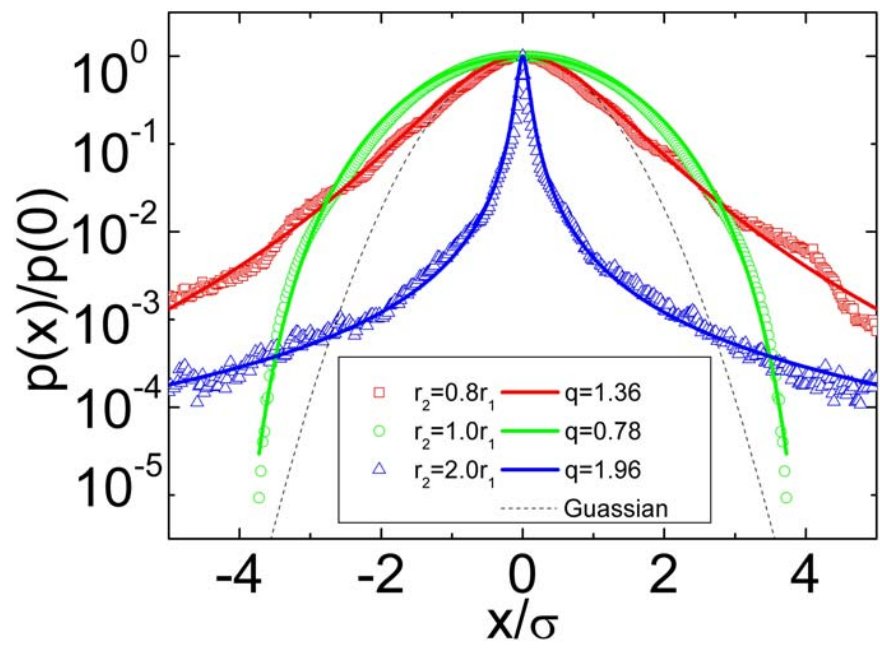

Fig.4

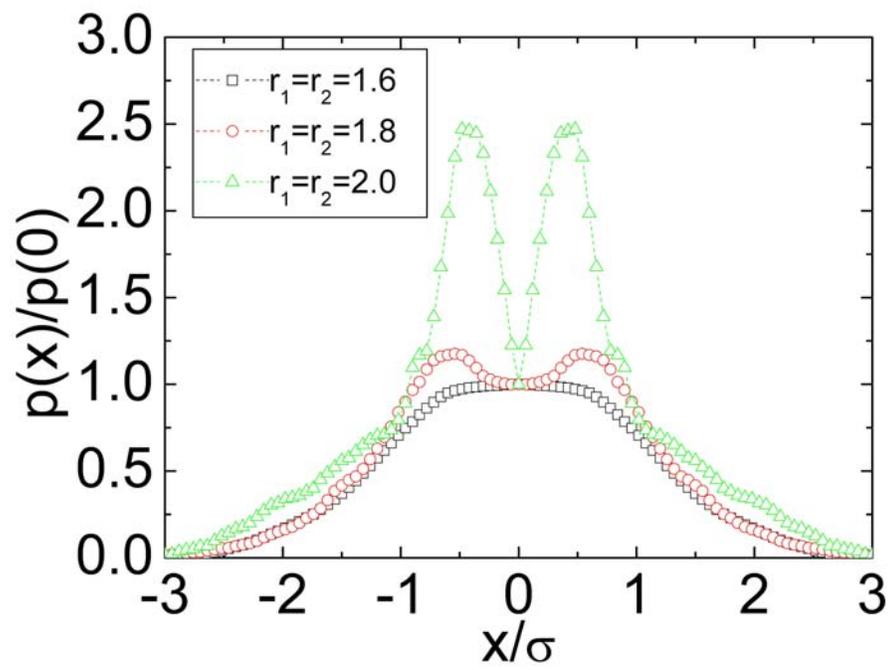

Fig.5